\documentclass[12pt]{article}

\ifx\pdfoutput\undefined
\usepackage[dvips,bookmarks]{hyperref}
\else
\usepackage{hyperref}
\fi
\hypersetup{colorlinks=false,bookmarksopen,bookmarksnumbered,citecolor=blue,
   pdfstartview=FitH}

\usepackage[dvips]{graphicx}
\usepackage{latexsym}
\usepackage{amssymb,amsfonts,amsmath}
\usepackage{graphicx}
\usepackage{indentfirst}
 \usepackage{bbm}
\usepackage{mathrsfs}

\usepackage{amsmath, amsthm,amssymb}
\usepackage{mathrsfs}
\usepackage{hyperref}
\usepackage{amsfonts}
\usepackage{dsfont}
\usepackage{slashed, tensor}
\usepackage{booktabs}
\usepackage{graphicx}
\usepackage{mathrsfs}

\parskip=\medskipamount

\usepackage{lipsum}       
\usepackage{xargs}            

\usepackage[colorinlistoftodos,prependcaption,textsize=tiny]{todonotes}
\newcommandx{\unsure}[2][1=]{\todo[linecolor=red,backgroundcolor=red!25,
bordercolor=red,#1]{#2}}
\newcommandx{\change}[2][1=]{\todo[linecolor=blue,
backgroundcolor=blue!25,bordercolor=blue,#1]{#2}}
\newcommandx{\STinfo}[1]{\todo[backgroundcolor=red!25,bordercolor=red,noline]{S.T.:#1}}
\newcommandx{\SKinfo}[1]{\todo[backgroundcolor=blue!25,bordercolor=blue,noline]{S.K.:#1}}

\newcommand {\cD}{{\cal D}}
\newcommand {\cE}{{\cal E}}

\newcommand {\cH}{{\cal H}}

\newcommand {\cK}{{\cal K}}
\newcommand {\cL}{{\cal L}}
\newcommand {\cM}{{\cal M}}
\newcommand {\cN}{{\cal N}}

\newcommand {\cT}{{\cal T}}

\newcommand {\cW}{{\cal W}}

\newcommand {\cZ}{{\cal Z}}

\def\a{\alpha}

\def\b{\beta}
\def\c{\chi}
\def\d{\delta}
\def\e{\epsilon}
\def\f{\phi}
\def\g{\gamma}
\def\G{\Gamma}

\def\j{\psi}

\def\l{\lambda}
\def\m{\mu}

\def\o{\omega}

\def\q{\theta}
\def\r{\rho}
\def\s{\sigma}
\def\t{\tau}

\def\x{\xi}
\def\z{\zeta}
\def\D{\Delta}
\def\F{\Phi}
\def\J{\Psi}
\def\L{\Lambda}
\def\O{\Omega}

\def\rd{{\rm d}}
\def\ri{{\rm i}}
\def\re{{\rm e}}

\newcommand{\ad}{{\dot{\alpha}}}                           
\newcommand{\bd}{{\dot{\beta}}}                            
\newcommand{\ve}{\varepsilon}                            
\newcommand{\cDB}{{\bar\cD}}                            

\newcommand{\pa}{\partial}                           
\newcommand{\hf}{\frac12}

\newcommand{\vf}{\varphi}

\newcommand{\be}{\begin{equation}}
\newcommand{\ee}{\end{equation}}
\newcommand{\bea}{\begin{eqnarray}}
\newcommand{\eea}{\end{eqnarray}}
\newcommand{\non}{\nonumber}
\newcommand{\ba}{\begin{array}}
\newcommand{\ea}{\end{array}}

\newcommand{\bm}[1]{\mbox{\boldmath$#1$}}

\def\double #1{#1{\hbox{\kern-2pt $#1$}}}

\newcommand{\gd}{{\dot\g}}
\newcommand{\dd}{{\dot\d}}

\newcommand{\sSL}{\mathsf{SL}}

\newcommand{\sU}{\mathsf{U}}

\newcommand{\bsubeq}{\begin{subequations}}
\newcommand{\esubeq}{\end{subequations}}

\numberwithin{equation}{section}

\begin{document}

\begin{titlepage}
\begin{flushright}
January, 2017 \\
\end{flushright}
\vspace{5mm}

\begin{center}
{\Large \bf 
Off-shell superconformal higher spin multiplets \\
in four dimensions 
}
\\ 
\end{center}

\begin{center}

{\bf Sergei M. Kuzenko${}^{a}$, Ruben Manvelyan${}^{b}$  and  Stefan Theisen${}^{c}$ 
} \\
\vspace{5mm}

\footnotesize{
${}^{a}${\it School of Physics and Astrophysics M013, The University of Western Australia\\
35 Stirling Highway, Crawley, W.A. 6009, Australia}}  
~\\
\vspace{2mm}
\footnotesize{
${}^{b}${\it Yerevan Physics Institute, Alikhanian Br. St. 2, 0036 Yerevan, Armenia}}
~\\
\vspace{2mm}
\footnotesize{
${}^{c}${\it Max-Planck-Institut f\"ur Gravitationsphysik, Albert-Einstein-Institut,\\
Am M\"uhlenberg 1, D-14476 Golm, Germany}
}
\vspace{2mm}
~\\

\end{center}

\begin{abstract}
\baselineskip=14pt
We formulate off-shell $\cN=1$  superconformal higher spin multiplets 
in four spacetime dimensions and briefly discuss their coupling to conformal supergravity. As an example, we explicitly work out the coupling of the superconformal gravitino multiplet to conformal supergravity. The corresponding action is super-Weyl invariant for arbitrary supergravity backgrounds. However, it is gauge invariant only if 
the supersymmetric Bach tensor vanishes. This is similar to linearised conformal supergravity in curved background.
\end{abstract}

\vfill

\vfill
\end{titlepage}

\allowdisplaybreaks

\section{Introduction}

The role of conformal field theories as cornerstones for the exploration 
of more general quantum field theories, which are connected to them via renormalization 
group flows, has been appreciated since a long time ago.  
Higher spin gauge theories \cite{Fronsdal1,Fronsdal2,FV,Vasiliev}
have an even longer history and have attracted 
considerable interest recently.  It is quite natural to combine the two symmetry principles and 
to study conformal higher spin theories \cite{FT}.  
A further symmetry which is compatible with conformal and 
higher spin symmetry is supersymmetry.  This leads to superconformal  higher spin theories, first advocated in  \cite{FL}, which 
are the main focus of this note. More specifically, we introduce
off-shell ${\cal N}=1$ superconformal higher spin multiplets in four dimensions
and analyse in some detail the problem of lifting such supermultiplets 
to curved backgrounds.
Our main technical tool, as far as the supersymmetry 
and supergravity aspects are concerned, is superspace and 
we refer to \cite{BK} for a thorough introduction to this formalism.  

We first study superconformal  higher spin  theories in flat superspace. In Sect. 2 we review superconformal transformations  
and the important notion of superconformal primaries.  In Sect. 3 we construct off-shell superconformal higher spin
multiplets: starting from prepotentials and their transformation laws under higher spin gauge transformations 
and under superconformal transformations, we construct field strengths and invariant actions. 
The two cases of half-integer and integer superspin as well as the superconformal gravitino 
multiplet have to be treated separately. The component fields of these multiplets are 
totally  symmetric traceless tensor and tensor-spinor fields. More general fields will be 
briefly discussed in the last part of Sect. 3.  In Sect. 4 we couple the superconformal higher spin multiplets to conformal supergravity where the notion of superconformal transformations is replaced by that of super-Weyl transformations.  
While super-Weyl invariance is easy to achieve, gauge invariance requires non-minimal couplings.  
We explicitly discuss the gravitino multiplet, but defer 
the general case to the future. Sect. 5 contains concluding comments, 
 including the  explicit expressions
for conserved higher spin current multiplets that correspond to the superconformal higher spin prepotentials. 
The main body of the paper is accompanied by two technical appendices. 
Appendix A contains those results concerning the Grimm-Wess-Zumino superspace geometry \cite{GWZ}, which are important for understanding 
the supergravity part of this paper. Appendix B contains the essential information 
about the super-Weyl transformations \cite{HT}.

There are different ways to describe $\cN=1$ conformal supergravity 
in superspace.\footnote{See \cite{FT} for a nice review of $\cN=1$ conformal supergravity and the complete list of references.}
The simplest option is to make use of the superspace geometry of \cite{GWZ}, 
which underlies the Wess-Zumino approach \cite{WZ} to the old minimal formulation for $\cN=1$ supergravity developed independently in \cite{old}.
Another option is to work with the $\sU(1)$ superspace proposed by Howe \cite{Howe}. 
Finally, one can make use of the so-called conformal superspace \cite{Butter}.
The three superspace approaches to $\cN=1$ conformal supergravity are 
equivalent, although each of them has certain advantages and disadvantages (see \cite{Butter} for a detailed discussion of the relationship between these formulations). 
In this paper we make use of the oldest and simplest superspace setting \cite{GWZ}.


\section{Superconformal transformations}

In this section we briefly recall the structure of $\cN=1$ superconformal transformations 
in Minkowski superspace ${\mathbb M}^{4|4}$, see \cite{BK} for more details. 
We denote by $z^A =(x^a, \q^\a , \bar \q_\ad)$ the Cartesian coordinates
for ${\mathbb M}^{4|4}$,  and use the notation $D_A =(\pa_a, D_\a , \bar D^\ad)$ 
for the  superspace covariant derivatives. 

Let $\x= \x^B  D_B =\x^b \pa_b + \x^\b D_\b +\bar \x_\bd \bar D^\bd$ 
be a real supervector field on ${\mathbb M}^{4|4}$.
It is called conformal Killing if it obeys the equation 
\bea
\Big[ \x + \hf K^{bc}[\x] M_{bc}\, , D_A \Big] +\d_{\s [\x]} D_A =0~, 
\label{master}
\eea
for some local Lorentz  ($K^{bc}[\x] $) and super-Weyl ($\s[\x]$) parameters. 
The super-Weyl transformation of the covariant derivatives is defined in  \eqref{superweyl}.
Choosing $A=\a$ and $A=\ad$ in \eqref{master} implies
that the spinor components of $\x^A$
as well as the parameters $K^{bc}[\x] $ and $\s[\x]$ are expressed in terms
of the vector components of $\x^A$:
\begin{subequations} 
\bea
\x^\a &=& -\frac{\ri }{8} \bar D_{\ad } \x^{\ad \a } ~, \qquad \bar D_\gd \x^\a =0~,\\
K_{\a\b}[\x]&=& D_{(\a} \x_{\b)}~, \qquad \bar D_\gd K_{\a\b}[\x] =0~, \\
\s [\x] &= & \frac{1}{3} ( D_\a \x^\a + 2 \bar D^\ad \bar \x_\ad )
~, \qquad \bar D_\gd \s[\x] =0
~.
\eea
\end{subequations}
The vector components of $\x^A$ obey the equations 
\bea
D_{(\a } \x_{\b) \bd} =0 \quad & \Longleftrightarrow &\quad 
\bar D_{(\ad} \x_{\bd) \b} =0 ~,
\eea
which imply
\bea
D^2 \x_{\b \bd} =0 \quad & \Longleftrightarrow &\quad 
\bar D^2 \x_{\bd \b} =0 ~,
\eea
as well as 
the ordinary conformal Killing equation
\bea
\pa_a \x_b + \pa_b \x_a = \hf
\eta_{ab} \pa_c \x^c
~.
\eea
A useful corollary of \eqref{master} with $A=\a$ is 
\bea
D_\g K^{\a\b} [\x] = \d_\g^{(\a} D^{\b)} \s[\x] \quad & \Longrightarrow &\quad 
D^2 \s[\x] =0
~.
\eea
Another consequence of \eqref{master} is 
\bea
\pa_a \s [\x] = \pa_a \bar \s [\x]  \quad & \Longrightarrow &\quad 
\pa_a D_\b  \s [\x]=0~.
\eea

The most general conformal Killing supervector field proves to be
\begin{subequations} \label{chiraltra}
\bea
 \x_+^{\ad\a} &=& a^{\ad\a}  +\hf (\s +{\bar \s})\, x_+^{\ad\a}
+{\bar K}^\ad{}_\bd  \,x_+^{\bd \a} +x_+^{\ad\b}K_\b{}^\a 
+x_+^{\ad \b} b_{\b \bd} x_+^{\bd \a} \non \\
&& \qquad +4{\rm i}\, {\bar \e}^\ad   \q^\a - 4 x_+^{\ad \b} \eta_\b \q^\a ~,
 \\
\x^\a &=& \e^\a + \big(\bar \s -\hf \s\big) \q^\a + \q^\b K_\b{}^\a 
+  +\q^\b b_{\b\bd}   x_+^{\bd \a}
-{\rm i}\,{\bar \eta}_\bd x_+^{\bd \a} +2 \q^2 \eta^\a~,
\eea
\end{subequations}
where we have introduced the complex four-vector
\be
\x^a_+ = \x^a + \frac{\rm i}{8} \x \s^a  {\bar \q}~, 
\qquad \bar \x^a =\x^a~,
\ee
along with the complex bosonic coordinates $x_+^a = x^a +\ri \q \s^a \bar \q$ 
of the chiral subspace of ${\mathbb M}^{4|4}$. 
The constant bosonic parameters in \eqref{chiraltra}
correspond to the spacetime translation ($a^{\ad \a}$), 
 Lorentz transformation ($K_\b{}^\a,~{\bar K}^{\ad}{}_{\bd})$,
 special conformal transformation
($ b_{\a \bd}$), and  combined scale and $R$-symmetry transformations 
($\s =\t -\frac{2}{3} \ri \vf$). The constant fermionic parameters in \eqref{chiraltra}
correspond to the $Q$-supersymmetry ($\e^\a$) and $S$-supersymmetry 
($\eta_\a$) transformations. The constant parameters $K_{\a\b}$ and $\s$ are obtained 
from $K_{\a\b}[\x]$ and $\s[\x]$, respectively,  by setting $z^A=0$.

A tensor superfield $\cT $ (with its  indices suppressed)
is said to be superconformal primary of weight $(p,q)$
if its superconformal transformation law is 
\bea
\d_\x \cT = \Big(\x + \hf K^{bc}[\x] M_{bc}\Big) \cT
+\Big(p \s[\x]+ q \bar \s[\x] \Big) \cT~,
\label{2.8}
\eea
for some parameters $p$ and $q$. 
The dimension of $\cT$ is $(p+q)$,
while $(p-q)$ is proportional to its $R$-symmetry charge.  
If $\cT$ is superconformal primary and chiral, $\bar \cD_\ad \cT =0$, 
then $\cT$ cannot possess dotted indices, i.e. $\bar M_{\ad\bd} \cT=0$, 
and it must hold that  $q=0$. In the chiral case, it suffices to say that 
$\cT$ is superconformal primary of dimension $p$.

Given a  real scalar $\cL$, which is superconformal primary of weight (1,1), 
\bea
\d_\x \cL = \x  \cL
+\big( \s[\x]+ \bar \s[\x] \big) \cL 
=\pa_a (\x^a \cL) -D_\a (\x^\a \cL)  -\bar D^\ad (\bar \x_\ad \cL)~,
\eea
 the functional
 \bea
 S= \int \rd^4x \rd^2 \q  \rd^2 \bar \q \, \cL 
\eea
 is  invariant under superconformal transformations.
Given a  chiral scalar $\cL_{\rm c}$,
which is superconformal primary of dimension $+3$, 
\bea
 \bar D_\ad \cL_{\rm c} =0~, \qquad \d_\x \cL_{\rm c} = \x  \cL_{\rm c}
+3  \s[\x] \cL_{\rm c}
=\pa_a (\x^a \cL_{\rm c}) -D_\a (\x^\a \cL_{\rm c})  ~,
\eea
 the  functional 
 \bea
 S_{\rm c}= \int \rd^4x \rd^2 \q  \, \cL_{\rm c} 
\eea
is invariant  under superconformal transformations.


\section{Off-shell superconformal multiplets in flat space}

In this section we introduce off-shell superconformal higher spin multiplets. 
We first consider the half-integer and integer superspin cases, and then give some 
generalisations of the constructions proposed. Strictly speaking, the notion of superspin is defined 
only for super-Poincar\'e multiplets. The rationale for our use of this name in the superconformal framework 
is that our superconformal multiplets will be 
described solely in terms of the gauge prepotentials corresponding to the off-shell 
massless higher spin multiplets constructed in \cite{KPS,KS93}.
Each of these massless multiplets also involves certain compensator 
superfields, in addition to the gauge prepotential.

\subsection{Half-integer superspin}

Let  $s$ be a positive integer. In the superspin-$(s+\hf)$ case, 
the conformal  prepotential $H_{\a(s)\ad(s)} \equiv H_{\a_1 \dots \a_s \ad_1 \dots \ad_s} $ is a real superfield, which is symmetric in its undotted indices and, independently, in its dotted indices.
The gauge transformation  law of  $H_{\a(s)\ad(s)} $ is 
\bea
 \d H_{\a_1 \dots \a_s \ad_1 \dots \ad_s} 
 = \bar D_{(\ad_1} \L_{\a_1 \dots \a_s \ad_2 \dots \ad_s )} 
- D_{(\a_1} \bar{\L}_{\a_2 \dots \a_s)\ad_1 \dots \ad_s} \ ,
\label{2.1}
\eea
with unconstrained gauge parameter $\L_{\a(s) \ad(s-1)}$. 

In the $s=1$ case, the transformation law \eqref{2.1} corresponds to linearised 
conformal supergravity \cite{FZ2}. The same transformation of $H_{\a\ad}$ 
occurs in all off-shell models for linearised $\cN=1$ supergravity, 
see \cite{BK} for a review. Such actions involve not only the gravitational superfield
\cite{FZ2,OS,Siegel} $H_{\a\ad}$, but also certain compensators. 
For $s>1$, the gauge transformation law \eqref{2.1} 
was introduced in \cite{KPS} in the framework of the (two dually equivalent) off-shell formulations for the massless superspin-$(s+\hf)$ multiplet.
The massless actions of \cite{KPS} involve not only 
the gauge prepotential $H_{\a(s) \ad(s)}$
but also certain compensators  (see \cite{BK} for a pedagogical review). 

The superconformal transformation law of $H_{\a(s) \ad (s)} $ is 
\bea
\d_\x H_{\a(s) \ad (s)} = \Big(\x + \hf K^{bc}[\x] M_{bc}\Big) H_{\a(s) \ad(s)}
-\frac{s}{2} \big(\s[\x]+\bar \s[\x] \big) H_{\a(s) \ad(s)}~.
\label{2.2}
\eea
This transformation law is uniquely determined if one requires 
both the gauge superfield $H_{\a(s) \ad(s)} $
and the gauge parameter $\L_{\a(s) \ad(s-1)}$ in \eqref{2.1} to be superconformal primary (see also \cite{KU,Park}).
It follows from \eqref{2.1} that the chiral symmetric spinor
\bea
\cW_{\a_1 \dots \a_{2s+1}} = -\frac{1}{4}\bar D^2 \pa_{(\a_1}{}^{\bd_1} \dots \pa_{\a_s}{}^{\bd_s}
D_{\a_{s+1}} H_{\a_{s+2} \dots \a_{2s+1} )\bd_1 \dots \bd_s} 
\label{2.3}
\eea
is gauge invariant \cite{KPS}.\footnote{The chiral superfield \eqref{2.3}  
is the only gauge-invariant field strength which remains non-vanishing on-shell 
in the supersymmetric higher spin theories introduced in \cite{KPS}. In a model independent framework of superfield representations, field strengths of the form \eqref{2.3} appeared in \cite{GGRS}.}
Our crucial observation is that $\cW_{\a(2s+1) }$ is superconformal primary of 
dimension $3/2$.
We conclude that the gauge-invariant action 
\bea
S_{s+\hf}= \int \rd^4x \rd^2 \q \, \cW^{\a_1 \dots \a_{2s+1}}\cW_{\a_1 \dots \a_{2s+1}} 
+ \int \rd^4x \rd^2 \bar \q \, \bar \cW_{\ad_1 \dots \ad_{2s+1}}
 \bar \cW^{\ad_1 \dots \ad_{2s+1}} 
 \label{2.5}
\eea
is superconformal. In the $s=1$ case, it coincides with the action for 
linearised conformal supergravity \cite{FZ2}.
One may check that 
\bea
 \int \rd^4x \rd^2 \q \, \cW^{\a_1 \dots \a_{2s+1}}\cW_{\a_1 \dots \a_{2s+1}} 
 =  \int \rd^4x \rd^2 \bar \q \, \bar \cW_{\ad_1 \dots \ad_{2s+1}}
 \bar \cW^{\ad_1 \dots \ad_{2s+1}} ~.
 \eea
 
We briefly comment on the component structure of the superconformal theory 
\eqref{2.5}.
The gauge parameter  $\bar D_{(\ad_1} \L_{\a_1 \dots \a_s \ad_2 \dots \ad_s )} $
in \eqref{2.1} may be represented as 
\bea 
\bar D_{(\ad_1} \L_{\a (s) \ad_2 \dots \ad_s )} (\q, \bar \q) = \re^{\ri \cH_0} 
\Big\{ g_{\a(s) \ad_1 \dots \ad_s} 
+ \ri \bar \q_{(\ad_1} \r_{\a(s) \ad_2 \dots \ad_s)}
&+& \ri \q^\b \c_{\b, \a(s) \ad_1 \dots \ad_s} \non \\
 + \q^2 v_{\a(s) \ad_1 \dots \ad_s} 
+\q^\b \bar \q_{(\ad_1} f_{\b, \a(s) \ad_2 \dots \ad_s)}
&+& \q^2 \bar \q_{(\ad_1} \o_{\a(s) \ad_2 \dots \ad_s)}\Big\} ~,
\label{2.7}
\eea
where $\cH_0 = \q \s^a \bar \q \pa_a$ and all component gauge parameters are complex. The parameters 
$\c_{\b, \a(s) \ad (s)} $ and $ f_{\b, \a(s) \ad (s-1)}$ 
transform in the representation 
$ \bf 2 \otimes  (\bm{2s+1})$ of $\sSL (2,{\mathbb C})$ with respect to their 
undotted indices. It follows from \eqref{2.1} and \eqref{2.7} that 
a Wess-Zumino gauge may be chosen of the form
\bea
H_{\a_1 \dots \a_s \ad_1 \dots \ad_s} (\q, \bar \q) 
&=& \q^\b \bar \q^\bd h_{(\b \a_1 \dots \a_s) (\bd \ad_1 \dots \ad_s) } 
+\bar \q^2 \q^\b \j_{(\b \a_1 \dots \a_s) \ad_1 \dots \ad_s} \non \\
&&- \q^2 \bar \q^\bd \bar \j_{ \a_1 \dots \a_s (\bd \ad_1 \dots \ad_s) } 
+\q^2 {\bar \q}^2 h_{\a_1 \dots \a_s  \ad_1 \dots \ad_s} ~,
\eea
where the bosonic fields $h_{\a(s+1) \ad(s+1)}$ and $h_{\a(s) \ad(s)}$ are real. 
The residual gauge freedom is generated by
\bea 
&&\bar D_{(\ad_1} \L_{\a (s) \ad_2 \dots \ad_s )} (\q, \bar \q) = \re^{\ri \cH_0} 
\Big\{ -\frac{\ri}{2}\, \z_{\a(s) \ad_1 \dots \ad_s} 
+\ri \bar \q_{(\ad_1} \r_{\a(s) \ad_2 \dots \ad_s)}
-\ri \q_{(\a_1} \bar \r_{\a_2 \dots \a_s) \ad_1 \dots \ad_s} \non \\
&&\qquad  +\frac{s}{s+1} \q^\b \bar \q_{(\ad_1} \pa_{(\b}{}^\gd 
\z_{\a_1 \dots \a_s ) \ad_1 \dots  \ad_{s-1}\gd }\non \\
&&\qquad - 
\hf \frac{s^2}{(s+1)^2} \q_{(\a_1} \bar \q_{(\ad_1} 
\pa^{\g\gd} \z_{\a_2 \dots \a_{s}) \g \ad_2 \dots \ad_{s}) \gd}
-2\ri\, \q_{(\a_1} \bar \q_{(\ad_1} \z_{\a_2 \dots \a_s) \ad_2 \dots \a_s)} \non \\
&& \qquad  - \frac{ s}{s+1} \q^2 \bar \q_{(\ad_1} 
\pa_{(\a_1 }{}^\gd \bar \r_{\a_2 \dots \a_s) \g \ad_2 \dots \ad_{s} )}~ 
\Big\}~,
\eea
where the bosonic parameters $\z_{\a(s)\ad(s)}$ and $\z_{\a(s-1) \ad(s-1)}$ are real.
The residual gauge transformations are:
\begin{subequations} \label{gauge310}
\bea
\d h_{\a_1 \dots \a_{s+1} \ad_1 \dots \ad_{s+1} } 
&=& \pa_{(\a_1 (\ad_1} \z_{\a_2\dots \a_{s+1}) \ad_2 \dots \ad_{s+1})}~, \\
\d h_{\a_1 \dots \a_{s} \ad_1 \dots \ad_{s} } 
&=& \pa_{(\a_1 (\ad_1} \z_{\a_2\dots \a_{s}) \ad_2 \dots \ad_{s})}~,\\
\d \j_{\a_1 \dots \a_{s+1} \ad_1 \dots \ad_{s} } 
&=& \pa_{(\a_1 (\ad_1} \r_{\a_2\dots \a_{s+1}) \ad_2 \dots \ad_{s})}~.
\eea
\end{subequations}
These transformation laws correspond to conformal higher spin fields \cite{FT}.

Reducing the actions \eqref{2.5} from superspace to components, for $s=1,2, \dots$, reproduces the conformal higher spin actions introduced 
by Fradkin and Tseytlin \cite{FT}.

\subsection{Integer superspin}

In the superspin-$s$ case, the superconformal multiplet is described in terms of an 
unconstrained prepotential
 $\J_{\a(s)\ad(s-1)} \equiv \J_{\a_1 \dots \a_s \ad_1 \dots \ad_{s-1}} $ 
and its conjugate $\bar \J_{\a(s-1)\ad(s)}$. The prepotential is symmetric in its undotted indices and, 
independently, in its dotted indices.
For $s>1$ the gauge freedom is
\bea
 \d \J_{\a_1 \dots \a_s \ad_1 \dots \ad_{s-1}} 
 = D_{(\a_1}\bar  \L_{\a_2 \dots \a_s)\ad_1 \dots \ad_{s-1}}
+  \bar D_{(\ad_1} \z_{\a_1 \dots \a_s \ad_2 \dots \ad_{s-1} )} 
 \ ,
\label{2.6}
\eea
with unconstrained gauge parameters $\bar \L_{\a(s-1) \ad(s-1)}$ 
and $\z_{\a(s) \ad(s-2)}$. The choice $s=1$ will be considered in 
subsection \ref{subsection3.3}.

As was shown in \cite{KS93}, 
the prepotential  $\J_{\a(s)\ad(s-1)} $ naturally originates within the so-called 
{\it longitudinal} formulation for the massless superspin-$s$ multiplet, 
which also makes use  of a real unconstrained compensator $H_{\a (s-1) \ad(s-1)}$.
The prepotential  $\J_{\a(s)\ad(s-1)} $ enters the action functional of \cite{KS93}
only via the {\it longitudinal linear} field strength 
$G_{\a(s) \ad(s)}:= \bar D_{(\ad_1 } \J_{\a(s) \ad_2 \dots \ad_s) }$, which is 
manifestly invariant under the $\z$-transformation \eqref{2.6}. 
On the other hand, in the non-superconformal case 
the gauge parameter $\Lambda$ is not arbitrary but instead has the form 
$\bar \L_{\a(s-1) \ad (s-1)} = D^\b L_{ (\a_1 \dots \a_{s-1} \b) \ad(s-1) } $, 
with $L_{\a(s) \ad(s-1)}$ unconstrained. 
This is not critical since one may always 
make $\bar \L_{\a(s-1) \ad (s-1)} $ unconstrained at the cost of introducing an 
additional compensator (in complete analogy with the massless gravi\-tino case 
considered in \cite{GS} and reviewed in \cite{BK}). For the massless superspin-$s$ multiplet, 
there exists another off-shell formulation which was 
constructed in \cite{KS93} and called {\it transverse}.
It is dual to the longitudinal one. 
It does not appear to be suitable to describe a superconformal multiplet.

The superconformal transformation law of $\J_{\a(s) \ad (s-1)} $ is 
postulated to be 
\bea
\d_\x \J_{\a(s) \ad (s-1)} = \Big(\x + \hf K^{bc}[\x] M_{bc}\Big) \J_{\a(s) \ad(s-1)}
-\frac{1}{2} \big( s\s[\x]+ (s-1) \bar \s[\x] \big) \J_{\a(s) \ad(s-1)}~.~~~~~~
\label{3.12}
\eea
It follows from \eqref{2.6} that the following chiral descendants of the prepotentials
\begin{subequations}\label{3.13}
\bea
\cW_{\a_1 \dots \a_{2s}} &=& -\frac{1}{4}\bar D^2 \pa_{(\a_1}{}^{\bd_1} \dots 
\pa_{\a_{s-1}}{}^{\bd_{s-1}}
D_{\a_{s}} \J_{\a_{s+1} \dots \a_{2s} )\bd_1 \dots \bd_{s-1}} ~,
\label{3.13a} 
\\
\cZ_{\a_1 \dots \a_{2s}} &=& -\frac{1}{4}\bar D^2 \pa_{(\a_1}{}^{\bd_1} \dots 
\pa_{\a_{s}}{}^{\bd_{s}}
D_{\a_{s+1}} \bar \J_{\a_{s+2} \dots \a_{2s} )\bd_1 \dots \bd_{s}} 
\label{3.13b}
\eea
\end{subequations}
are gauge invariant.\footnote{The field strength \eqref{3.13a} was introduced in \cite{KS93}.}  
As before, the crucial observation is that the field strengths $\cW_{\a(2s) } $ and $\cZ_{\a(2s) }$ are superconformal 
primaries of dimension 1 and 2, respectively.  
This allows us to construct a superconformal and gauge-invariant action
\bea
S_{s}=\ri  \int \rd^4x \rd^2 \q \, \cW^{\a_1 \dots \a_{2s}}\cZ_{\a_1 \dots \a_{2s}} 
-\ri  \int \rd^4x \rd^2 \bar \q \, \bar \cW_{\ad_1 \dots \ad_{2s}}
 \bar \cZ^{\ad_1 \dots \ad_{2s}} ~.
 \label{action315}
\eea
One checks that 
\bea
 \int \rd^4x \rd^2 \q \, \cW^{\a_1 \dots \a_{2s}}\cZ_{\a_1 \dots \a_{2s}} 
 + \int \rd^4x \rd^2 \bar \q \, \bar \cW_{\ad_1 \dots \ad_{2s}}
 \bar \cZ^{\ad_1 \dots \ad_{2s}} =0~.
 \eea

We now comment on the component structure of 
\eqref{action315}. One may choose a Wess-Zumino gauge of the form
\bea
\J_{\a_1 \dots \a_s \ad_1 \dots \ad_{s-1}} (\q, \bar \q) 
&=& \q^\b \bar \q^\bd \j_{(\b \a_1 \dots \a_s) (\bd \ad_1 \dots \ad_{s-1}) } 
+\bar \q^2 \q^\b B_{(\b \a_1 \dots \a_s) \ad_1 \dots \ad_{s-1}} \non \\
&&- \q^2 \bar \q^\bd {\bm h}_{ \a_1 \dots \a_s (\bd \ad_1 \dots \ad_{s-1}) } 
+\q^2 {\bar \q}^2 \j_{\a_1 \dots \a_s  \ad_1 \dots \ad_{s-1}} ~,
\eea
where the bosonic fields ${\bm h}_{\a (s) \ad (s) }$ and $B_{\a(s+1) \ad(s-1)}$ 
are complex. In the Wess-Zumino gauge chosen, the bosonic fields ${\bm h}_{\a (s) \ad (s) }$ and $B_{\a(s+1) \ad(s-1)}$ and 
the fermionic fields $\j_{\a(s+1) \ad(s)}$ and $\j_{\a(s) \ad(s-1)}$
are defined modulo gauge freedom
of the type \eqref{gauge310}.\footnote{The case $s=1$ is not considered here. 
Its special feature  is that  $B_{\a\b}$ is not a gauge field. } 

More specifically, the field $B_{\a(s+1) \ad(s-1)}$ belongs to a more general family
of conformal fields than those described by the gauge transformation laws \eqref{gauge310}.
The point is that one may consider conformal higher spin fields 
$\f_{\a(m) \ad(n)}$, where $m$ and $n$ are integers such that $m > n > 0$. 
Since $m\neq n$, the field $\f_{\a(m) \ad(n)}$ is complex.
Postulating the gauge transformation law 
\bea
\d \f_{\a_1 \dots \a_m \ad_1 \dots \ad_n} 
= \pa_{(\a_1 (\ad_1 } \l_{ \a_2 \dots \a_m) \ad_2 \dots \ad_n)}
\eea
and requiring both the field $\f_{\a(m) \ad(n)}$ and the gauge parameter 
$\l_{\a(m-1) \ad (n-1)}$ to be primary, the dimension of $\f_{\a(m) \ad(n)}$
is fixed to be equal to $2- \hf (m+n)$. 
We can define two gauge-invariant field strengths
\begin{subequations}
\bea
\hat{C}_{\a_1 \dots \a_{m+n}} &=& \pa_{(\a_1}{}^{\bd_1} \dots \pa_{\a_n}{}^{\bd_n}
\f_{\a_{n+1} \dots \a_{m+n} ) \bd_1 \dots \bd_n}~,\\
\check{C}_{\a_1 \dots \a_{m+n}} &=& \pa_{(\a_1}{}^{\bd_1} \dots \pa_{\a_m}{}^{\bd_m}
\bar \f_{\a_{m+1} \dots \a_{m+n} ) \bd_1 \dots \bd_m}~.
\eea
\end{subequations}
They are conformal primaries of dimension
$2- \hf (n-m)$ and $2- \hf (m-n)$, respectively. 
In terms of those we can write a gauge-invariant conformal action 
\bea
S= \ri^{m+n} \int \rd^4 x \,\hat{C}^{\a_1 \dots \a_{m+n}} \check{C}_{\a_1 \dots \a_{m+n}} 
+{\rm c.c.}
\eea

\subsection{Superconformal gravitino multiplet} \label{subsection3.3}

In the $s=1$ case, the gauge transformation law \eqref{2.6} has to be 
replaced with 
\bea
\d \J_\a = D_\a \bar \L + \z_\a ~, \qquad \bar D_\bd \z_\a =0~.
\label{3.21}
\eea
This gauge transformation was given in Ref. \cite{GS}, which
proposed the off-shell formulation for the massless gravitino multiplet 
in terms of the gauge 
spinor prepotential $\J_\a$ in conjunction with two compensators,  
an unconstrained real scalar and a chiral scalar. 

The prepotential $\J_\a$ is required to be superconformal primary of weight 
$(-1, 0)$,
which is a special case of \eqref{3.12}. The superconformal primary superfields
\eqref{3.13} for $s=1$ are obviously invariant under the gauge transformations
\eqref{3.21}.


\subsection{Generalisations}

Given two integers $m>n>0$, 
we introduce a gauge prepotential
 $\F_{\a(m)\ad(n)} \equiv \F_{\a_1 \dots \a_m \ad_1 \dots \ad_n} $ 
and its conjugate $\bar \F_{\a(n)\ad(m)}$. 
The gauge transformation of $\F_{\a(m) \a(n) }$ is postulated to be 
\bea
 \d \F_{\a_1 \dots \a_m \ad_1 \dots \ad_{n}} 
 = D_{(\a_1}\bar  \L_{\a_2 \dots \a_m)\ad_1 \dots \ad_{n}}
+  \bar D_{(\ad_1} \z_{\a_1 \dots \a_m \ad_2 \dots \ad_{n} )}  \ ,
\eea
with unconstrained gauge parameters $\bar \L_{\a (m-1) \ad (n)} $ and $\z_{\a(m)\ad(n-1)}$. 
The superconformal transformation law of $\F_{\a(m) \ad (n)} $ is 
\bea
\d_\x \F_{\a(m) \ad (n)} = \Big(\x + \hf K^{bc}[\x] M_{bc}\Big) \F_{\a(m) \ad(n)}
-\frac{1}{2} \big( m\s[\x]+ n \bar \s[\x] \big) \F_{\a(m) \ad(n)}~.~~~~~~
\eea
Given $\Phi$, we can define two gauge-invariant chiral field strengths 
\begin{subequations}
\bea
{\mathbb W}_{\a_1 \dots \a_{m+n+1}} &=& -\frac{1}{4}\bar D^2 \pa_{(\a_1}{}^{\bd_1} \cdots 
\pa_{\a_{n}}{}^{\bd_{n}}
D_{\a_{n+1}} \F_{\a_{n+2} \dots \a_{m+n+1} )\bd_1 \dots \bd_{n}} ~,\\
{\mathbb Z}_{\a_1 \dots \a_{m+n+1}} &=& -\frac{1}{4}\bar D^2 \pa_{(\a_1}{}^{\bd_1} 
\cdots  \pa_{\a_{m}}{}^{\bd_{m}}
D_{\a_{m+1}} \bar \F_{\a_{m+2} \dots \a_{m+n+1} )\bd_1 \dots \bd_{m}}~.
\eea
\end{subequations}
which are
superconformal primaries of dimension $\hf (3+n-m)$ and 
$\hf (3+m-n)$, respectively. 
Therefore, the  following gauge-invariant action 
\bea
S=\ri^{m+n}  \int \rd^4x \rd^2 \q \, {\mathbb W}^{\a_1 \dots \a_{m+n+1}}
{\mathbb Z}_{\a_1 \dots \a_{m+n+1}} +{\rm c.c.} 
\eea
is superconformal. 


\section{Off-shell superconformal multiplets in supergravity}

We now turn to exploring whether the superconformal higher spin multiplets introduced
in the previous section may be consistently lifted to curved superspace backgrounds. 

\subsection{General considerations}
 
Just as in the non-supersymmetric setting, where conformal invariance in Minkowski space 
is replaced by Weyl invariance, in a curved background geometry, superconformal invariance  
is replaced by super-Weyl invariance.
In other words,  super-Weyl invariance in curved superspace implies superconformal 
invariance in Minkowski superspace. 

A tensor superfield $\cT $ (with its  indices suppressed)
is said to be super-Weyl primary of weight $(p,q)$
if its super-Weyl transformation law is 
\bea
\d_\s \cT =\big(p\, \s + q\, \bar \s \big) \cT~,
\label{B.3}
\eea
for some parameters $p$ and $q$.
Similar to the rigid supersymmetric case
\eqref{2.8}, we will refer to $(p+q)$ as the dimension of $\cT$.
Given a covariantly chiral tensor superfield $\cT$
defined on a general supergravity 
background, $\bar \cD_\ad \cT=0$,
it may carry only undotted indices,
$\bar M_{\ad \bd} \cT=0$, as a consequence of 
\eqref{A.3b}. If $\cT$ is covariantly chiral and  super-Weyl primary, eq. \eqref{B.3}, 
then $q=0$. An example is provided by the super-Weyl tensor $W_{\a\b\g}$
with the transformation law \eqref{s-WeylW}. In Appendix B we also collect
the transformation properties of various other geometric quantities under super-Weyl transformations. 

 As reviewed in Appendix \ref{AppendixA}, 
 the curved superspace geometry of \cite{GWZ}
  does not possess torsion tensors 
 of dimensions 1/2. This means that  the gauge transformation \eqref{2.1}
is uniquely extended to curved superspace as
\bea
 \d H_{\a_1 \dots \a_s \ad_1 \dots \ad_s} 
 = \bar \cD_{(\ad_1} \L_{\a_1 \dots \a_s \ad_2 \dots \ad_s )} 
- \cD_{(\a_1} \bar{\L}_{\a_2 \dots \a_s)\ad_1 \dots \ad_s} \ .
\label{4.1}
\eea
It is compatible with the following super-Weyl transformation of 
the prepotential: 
\bea
\d_\s H_{\a(s) \ad (s)} = -\frac{s}{2} \big(\s +\bar \s  \big) H_{\a(s) \ad(s)}~.
\eea

The chiral field strength \eqref{2.3} may uniquely be lifted to curved superspace 
as a covariantly chiral superfield of the general form
\begin{subequations}
\bea
\cW_{\a (2s+1)} &=& 
-\frac{1}{4}(\bar \cD^2 -4R)\Big\{ 
\cD_{(\a_1}{}^{\bd_1} \cdots \cD_{\a_s}{}^{\bd_s}
\cD_{\a_{s+1}} H_{\a_{s+2} \dots \a_{2s+1} )\bd_1 \dots \bd_s} + \dots \Big\}~, 
\label{4.3a}\\
\noalign{\vskip.2cm}
& & \bar \cD_\bd \cW_{\a (2s+1)} = 0~,
\eea
\end{subequations}
with the super-Weyl transformation law 
\bea
\d_\s  \cW_{\a (2s+1)} =\frac{3}{2} \s \cW_{\a (2s+1)}~.
\label{4.4}
\eea
The ellipsis in \eqref{4.3a} stands for terms involving the super-Ricci tensor 
$G_{\a\ad}$ and its covariant derivatives. Such terms can always be found. 
A systematic construction is to start with  conformal superspace \cite{Butter},
where $G_{\a\ad}$ appears as a connection, and then 
to implement the so-called de-gauging procedure in order  to arrive 
at the ordinary curved superspace geometry of \cite{GWZ}.\footnote{In conformal superspace,
the required primary chiral field strength  $\cW_{\a (2s+1)} $ has a minimal form
$\cW_{\a (2s+1)} = -\frac{1}{4}\bar \nabla^2 
\nabla_{(\a_1}{}^{\bd_1} \cdots \nabla_{\a_s}{}^{\bd_s}
\nabla_{\a_{s+1}} H_{\a_{s+2} \dots \a_{2s+1} )\bd_1 \dots \bd_s}$,
where $\nabla_A = (\nabla_a , \nabla_\a , \bar \nabla^\ad)$ denotes 
the corresponding covariant derivatives \cite{Butter}.} 
Details of the construction will be given elsewhere, 
but examples of the complete superfields for $s=1$ and $s=2$ are given below 
in \eqref{4.6}  and \eqref{499}, respectively. Two  observations, 
which are crucial for this construction, are that the descendant
$A_{\a(s+1) \bd(s)} :=\cD_{(\a_{s+1}} H_{\a_{1} \dots \a_{s} )\bd_1 \dots \bd_s} $ 
is super-Weyl primary and obeys the constraint 
$\cD_{(\a_1} A_{\a_2 \dots \a_{s+2} ) \bd(s)} =0$.

We may now consider a minimal 
extension of \eqref{2.5} to curved superspace given by 
\bea
 \int \rd^4x \rd^2 \q \, \cE \,\cW^{\a_1 \dots \a_{2s+1}}\cW_{\a_1 \dots \a_{2s+1}} 
+{\rm c.c.}~,
\label{4.5}
 \eea
where $\cE$  is the chiral integration measure.
It follows from \eqref{4.4} that this functional  is super-Weyl invariant.
However, for non-vanishing  background super-Weyl tensor, $W_{\a\b\g} \neq 0$,
the field strength $\cW_{\a (2s+1)} $ and, therefore,  the action \eqref{4.5}
are not gauge invariant. In general, the gauge variation 
$\d_\L \cW_{\a (2s+1)} $ is proportional 
to the background super-Weyl tensor $W_{\a\b\g}$, its conjugate
 $\bar W_{\ad\bd \gd}$ 
and their covariant derivatives.\footnote{For instance, in the $s=1$ case the variation
$\d_\L \cW_{\a (3)} $ is given by \eqref{4.8}.}
The action \eqref{4.5} needs to be completed to include non-minimal terms
which contain $W_{\a\b\g}$, $\bar W_{\ad\bd \gd}$ 
and their covariant derivatives. An example will be given 
in section 4.3, where we discuss the gravitino supermultiplet. 

Let us first discuss the simplest case of $\cW_{\a(2s+1)}$, $s=1$, which
is linearised conformal supergravity. 
The linearised super-Weyl tensor is
\bea
\cW_{\a\b\g} = -\frac{1}{4} (\bar \cD^2 - 4R) 
\Big\{(\cD_{(\a}{}^\gd +\ri G_{(\a}{}^\gd) \cD_\b 
H_{\g)\gd} \Big\}~,
\label{4.6}
\eea
 modulo normalisation.
It varies homogeneously under the super-Weyl transformation,
in accordance with \eqref{4.4}.
However, $\cW_{\a\b\g} $ is not invariant under the gauge transformation 
\eqref{4.1} with $s=1$. One may check that 
\bea
\d_\L  \cW_{\a\b\g} = \frac{\ri}{2} (\bar \cD^2 - 4R) \Big[
\big(\cD^\d W_{\d(\a\b} \big) \L_{\g)} 
-\cD_{(\a} \big(W_{\b\g) \d} \L^\d \big)\Big]~.
\label{4.8}
\eea
The important point is that each term in $\d_\L  \cW_{\a\b\g} $ involves either 
the background super-Weyl tensor or its covariant derivative. The variation vanishes if
the background superspace is conformally flat, $W_{\a\b\g}=0$.
In this case the functional \eqref{4.5} is the required 
superconformal gauge-invariant action. Here `superconformal' means that 
the action is invariant under arbitrary superconformal isometries of the background 
superspace.

As another example of $\cW_{\a(2s+1)}$, we consider the case $s=2$. 
The field strength $\cW_{\a(5)}$ is uniquely determined to be 
\bea
\cW_{\a_1 \dots \a_5} = -\frac{1}{4} (\bar \cD^2 - 4R) 
\Big\{ \cD_{(\a_1}{}^{\bd_1} \cD_{\a_2}{}^{\bd_2} 
&+& 3\ri G_{(\a_1}{}^{\bd_1} \cD_{\a_2}{}^{\bd_2} 
-2 G_{(\a_1}{}^{\bd_1} G_{\a_2}{}^{\bd_2} \non \\
 -\frac{1}{4} ([\cD_{(\a_1} , \bar \cD^{\bd_1} ] G_{\a_2}{}^{\bd_2})
&+& \frac{3}{2} \ri (\cD_{(\a_1}{}^{\bd_1} G_{\a_2}{}^{\bd_2} ) \Big\}
\cD_{\a_3} H_{\a_4 \a_5 ) \bd_1 \bd_2}~.
\label{499}
\eea
It is a tedious exercise to check that $\cW_{\a(5)}$  is super-Weyl primary.

In the case of anti-de Sitter superspace ${\rm AdS}^{4|4}$ \cite{Keck,Zumino77,IS}
specified by 
\bea
W_{\a\b\g} =0~, \qquad G_{\a\ad}=0~,\qquad R\neq 0~,
\eea
the gauge-invariant chiral field strength $\cW_{\a(2s+1)}$ was found in \cite{KS94}.
It is
\bea
\cW_{\a (2s+1)} &=& -\frac{1}{4}(\bar \cD^2 -4R)\cD_{(\a_1}{}^{\bd_1} \cdots \cD_{\a_s}{}^{\bd_s}
\cD_{\a_{s+1}} H_{\a_{s+2} \dots \a_{2s+1} )\bd_1 \dots \bd_s} ~.
\eea
The curved-superspace extension of the gauge transformation \eqref{2.6} is
\bea
 \d \J_{\a_1 \dots \a_s \ad_1 \dots \ad_{s-1}} 
 = \cD_{(\a_1}\bar  \L_{\a_2 \dots \a_s)\ad_1 \dots \ad_{s-1}}
+  \bar \cD_{(\ad_1} \z_{\a_1 \dots \a_s \ad_2 \dots \ad_{s-1} )}  \ .
\eea
It is compatible with the following super-Weyl transformation of the prepotential 
\bea
\d_\s \J_{\a(s) \ad (s-1)} =-\frac{1}{2} \big( s\s  + (s-1) \bar \s \big) \J_{\a(s) \ad(s-1)}~.
\eea

\subsection{Superconformal gravitino multiplet} 

Our next example is the superconformal gravitino multiplet.  
It is characterised by the gauge freedom
\bea
\d \J_\a = \cD_\a \bar \L + \z_\a ~, \qquad \bar \cD_\bd \z_\a =0~, 
\label{3.10}
\eea
and the super-Weyl transformation
\bea
\d_\s \J_{\a} =-\frac{1}{2} \s  \J_{\a}~.
\eea
The following covariantly chiral field strengths 
\begin{subequations}
\bea
\cW_{\a\b}&=& -\frac{1}{4} (\bar \cD^2 -4R) \cD_{(\a} \J_{\b)}~,\\
\cZ_{\a\b} &=& -\frac{1}{4} (\bar \cD^2 -4R) \Big[ (\cD_{(\a}{}^\ad +\ri G_{(\a}{}^\ad ) 
\cD_{\b)} \bar \J_\ad\Big] 
\eea
\end{subequations}
are super-Weyl primary of dimension $+1$ and $+2$, respectively.
These superfields are not invariant under the gauge transformations \eqref{3.10}.
One finds the following non-vanishing variations of $\cW_{\a\b}$ and $\cZ_{\a\b}$:
\begin{subequations}
\bea
\d_\z \cW_{\a\b} &=& 2 W_{\a\b \g} \z^\g~,\\
\d_\L \cZ_{\a\b} &=& \frac{\ri}{2} W_{\a\b\g} (\bar \cD^2 -4R ) \cD^\g \L 
 + \frac{\ri}{2} (\bar \cD^2 -4R ) \Big\{ \L \cD^\g W_{\a\b\g} \Big\}  ~.
\eea
\end{subequations}

Consider the action 
\bea
S_{\rm GM}&=&\ri  \int \rd^4x \rd^2 \q \, \cE\, \cW^{\a \b}\cZ_{\a \b} 
- 2\ri \int \rd^4x \rd^2 \q  \rd^2 \bar \q \, E\, W^{\a\b\g} \J_\a  (\cD_{\b \bd} + \ri G_{\b\bd}) 
\cD_\g \bar \J^\bd \non \\
&& + \int \rd^4x \rd^2 \q  \rd^2 \bar \q \, E\, (\cD_\a W^{\a\b\g})
(\bar \cD_\bd \J_\b)  \cD_\g \bar \J^\bd ~+~{\rm c.c.} 
\label{3.15}
\eea
Here $\cE$ and $E$ denote the chiral measure and the full superspace measure, respectively.
The action $S_{\rm GM}$ is super-Weyl invariant, 
\bea
\d_\s S_{\rm GM}=0~.
\eea
The second and third terms on the right of \eqref{3.15} are fixed by requiring 
$S_{\rm GM}$ to be invariant under the $\z$-transformation \eqref{3.10}, 
\bea
\d_\z S_{\rm GM} =0~.
\label{3.17}
\eea
Finally, a lengthy calculation gives 
\bea
\d_\L S_{\rm GM}= 2 \int \rd^4x \rd^2 \q  \rd^2 \bar \q \, E\,B^{\a\ad}
(\J_\a \bar \cD_\ad \L + \L \bar \cD_\ad \J_\a) +{\rm c.c.}
\label{3.18}
\eea
Here $B^{\a\ad}$ denotes the $\cN=1$ supersymmetric extension of the Bach tensor, 
\bea
B^\a{}_\ad &=&  \ri \cD_{\b \ad} \cD_\g W^{\a\b\g}
+ (\cD_\b G_{\g \ad}) W^{\a\b\g}+G_{\b \ad} \cD_\g W^{\a\b\g} \non \\
&=&\ri \cD_{\a \bd} \bar \cD_\gd \bar W^{\ad\bd\gd}
- (\bar \cD_\bd G_{\a \gd}) \bar W^{\ad\bd\gd} - G_{\a \bd} \bar \cD_\gd \bar W^{\ad\bd\gd}~,
\eea
with the super-Weyl transformation 
\bea
\d_\s B_{\a\ad} = \frac{3}{2} (\s+\bar \s) B_{\a\ad}~.
\eea
One can rewrite $B_{\a\ad}$ is a manifestly real form \cite{BK,BK88}
\bea
B_{\a\ad} &=& - \cD^b \cD_b G_{\a\ad} - W_\a{}^{\b\g} \cD_\b G_{\g \ad} 
+\bar W_\ad{}^{\bd \gd} \bar \cD_\bd G_{\a\gd} \\
&& +\frac{1}{4}\Big( (\cD^\b R) \cD_\b +(\bar \cD_\bd \bar R) \bar \cD^\bd \Big) 
G_{\a\ad} -(\bar \cD_\ad G^b)\cD_\a G_b - 3R \bar R G_{\a\ad} \non \\
&& +\frac{1}{8}G_{\a\ad} (\bar \cD^2 \bar R + \cD^2 R) 
+\frac{\ri}{4} \cD_{\a\ad} (\bar \cD^2 \bar R - \cD^2 R)~.
\eea
We recall that the super-Bach tensor  may be introduced 
(see \cite{BK,BK88} for the technical details)
as a functional derivative of the conformal 
supergravity action  \cite{Siegel78,Zumino},
\bea 
I_{\rm CSG} =  \int \rd^4x\, \rd^2\q\, \cE\,  W^{\a\b \g}W_{\a\b\g} 
+{\rm c.c.} ~,
\label{4.24}
\eea
with respect to the gravitational superfield, specifically
\bea
\d  \int \rd^4x \rd^2 \q \, \cE\, W^{\a \b \g}W_{\a\b\g } =
 \int \rd^4x \rd^2 \q  \rd^2 \bar \q \, E\, \D H^{\a\ad} B_{\a\ad}~,
 \eea
with $\D H^{\a\ad} $ the covariantised variation of the gravitational superfield
defined in \cite{GrisaruSiegel}. The super-Bach tensor obeys the conservation equation 
\bea
\cD^\a B_{\a\ad}=0 \quad & \Longleftrightarrow &\quad  \bar \cD^\ad B_{\a\ad} =0~,
\eea
which expresses the gauge invariance of the conformal supergravity action. 

It follows from \eqref{3.17} and \eqref{3.18} that the action \eqref{3.15} 
is gauge invariant if the background super-Bach tensor is equal to zero, 
\bea
B_{\a\ad}=0~.
\label{4.28}
\eea
This holds, e.g.,  for all Einstein superspaces, which are characterised by
\bea 
G_{\a \ad}=0 \quad \Longrightarrow \quad R ={\rm const}~.
\eea


\subsection{Linearised conformal supergravity}

The condition \eqref{4.28} is also required to define
an off-shell superconformal multiplet of superspin 3/2 in curved superspace.
The point is that \eqref{4.28} is the equation of motion for conformal 
supergravity, since varying the action \eqref{4.24} 
with respect to the gravitational superfield gives\footnote{The two terms in the 
right-hand side of \eqref{4.24} differ by a total derivative 
related to the Pontryagin invariant \cite{BK,BK88}.}
\bea
\d  S_{\rm CSG} =
2 \int \rd^4x \rd^2 \q  \rd^2 \bar \q \, E\, \D H^{\a\ad} B_{\a\ad}~.
 \eea
The gauge-invariant action for the superconformal superspin-$\frac{3}{2}$ multiplet
in curved background is obtained by 
linearising the conformal supergravity action \eqref{4.24}
around its arbitrary stationary point,   $B_{\a\ad}=0$.
In accordance with  \cite{GrisaruSiegel} (see also \cite{BK} for a review), 
the linearised gauge transformation of the prepotential is given by \eqref{4.1} with $s=1$. The linearised conformal supergravity action 
is automatically invariant under the gauge and super-Weyl transformations. 
Its explicit structure will be described elsewhere. 


\section{Concluding comments} 

In this paper we constructed the off-shell $\cN=1$ superconformal higher spin multiplets in four dimensions\footnote{In three dimensions, the off-shell superconformal higher spin multiplets have recently been described  in \cite{KTs}
and \cite{KO} for  the $\cN=1$  and $\cN=2$ cases, respectively.} 
and also sketched the general scheme of coupling such multiplets to conformal 
supergravity. Our work opens two new approaches to interacting conformal higher spin
theories. Firstly, every conformal higher spin field may be embedded into an off-shell 
superconformal multiplet
(the latter actually contains several bosonic and fermionic conformal fields).
 Instead of trying to couple the original conformal field to gravity, we can look for a consistent interaction of the superconformal multiplet with conformal supergravity. 
 Since the gravitational field belongs to the conformal supergravity multiplet
 (also known as the Weyl multiplet), this will automatically lead to a consistent 
 coupling of the component conformal fields to gravity. 

The second avenue to explore is a superfield extension of the effective action 
approach to conformal higher spin fields advocated in 
\cite{Tseytlin2002,Segal,BJM}.
One may start with a free massless chiral scalar superfield $\F$, $\bar D_\ad \F=0$, 
and couple it to an infinite tower of background superconformal higher spin prepotentials $H_{ \a (s) \ad (s) }$ (source superfields)
by the rule
\bea
S[\F, \bar \F; H]=  \int \rd^4x \rd^2 \q  \rd^2 \bar \q \, \Big\{ 
\F \bar \F  + \sum_{s=1}^{\infty} H^{ \a (s) \ad (s) } J_{ \a (s) \ad (s)} \Big\}~,
\quad \bar J_{\a (s) \ad(s)} = J_{\a(s) \ad(s)}~.~~~~
\label{5.1}
\eea
Here $J_{ \a (s) \ad (s) }$ denotes a composite primary superfield, which describes a 
conserved current multiplet when $\F$ is on-shell. Then it is natural to consider 
the generating functional for correlation functions of these conserved higher spin 
supercurrents defined by
\bea
\re^{ \ri \G[H] } = \int \cD \F \cD \bar \F \,\re^{ \ri S[\F, \bar \F; H]}~.
\eea
Similar to the non-supersymmetric analysis of \cite{BJM2}, one
may show that the action \eqref{5.1}, 
properly deformed by terms nonlinear in  $H^{ \a (s) \ad (s) }  $, 
 has an exact non-Abelian gauge symmetry
(associated with the conformal higher-spin superalgebra $\mathsf{shsc}^\infty(4|1)$ 
described in \cite{FL})
which reduces to \eqref{4.1} at lowest level in the superfields 
$H_{ \a (s) \ad (s) }$, $\F$ and $\bar \F$.
Here we restrict our discussion 
to giving the explicit expressions for $J_{ \a (s) \ad (s) }$.

We turn to describing the structure of the conserved current multiplets 
$J_{\a (s) \ad(s)}$.  In order for the source term
\bea
S^{(s+\hf)}_{\rm source}=\int \rd^4x \rd^2 \q  \rd^2 \bar \q \, H^{ \a (s) \ad (s) } J_{ \a (s) \ad (s) }
\label{5.2}
\eea
to be invariant under the superconformal
transformations, the superfield $J_{\a(s) \ad(s)}$ must be 
superconformal primary of weight $(1+ \frac{s}{2},1+ \frac{s}{2})$.
 In order for $S^{(s+\hf)}_{\rm source}$ to be invariant under the gauge transformations 
 \eqref{2.1}, $J_{\a(s) \ad(s)}$  must obey the conservation equations
 \bea
 D^\b J_{\b \a_1 \dots \a_{s-1} \ad_1 \dots \ad_s }=0 ~,\qquad
 \bar D^\bd J_{\a_1 \dots \a_s \bd \ad_1 \dots \ad_{s-1}}=0~.
 \label{5.3}
 \eea
The $s=1$ case corresponds to the superconformal version 
\cite{FZ2} of the Ferrara-Zumino  supercurrent \cite{FZ}. 
Extension to the other cases $s>1$ was given in \cite{HST} (building on 
\cite{Siegel:1980bp}).
The authors of \cite{HST} also postulated the prepotential 
$H_{ \a (s) \ad (s) }$ as the source to generate the Noether coupling \eqref{5.2}, 
as well as  the gauge transformation \eqref{2.1} as the transformation
of $H_{ \a (s) \ad (s) }$ which leaves \eqref{5.2} invariant. 
However, no higher spin extensions of  linearised conformal supergravity were given.

Consider a free on-shell massless chiral scalar $\F$, 
\bea
\bar D_\ad \F =0~, \qquad D^2 \F =0~.
\eea
which is superconformal primary of dimension $+1$.
By analogy with the construction of \cite{CDT},
the conserved current multiplets $J_{\a(s) \ad (s)} $, with $s=1, 2, \dots,$
can be obtained as unique composites of $\F$ and $ \bar \F$ of the form 
\bea
J_{\a(s) \ad (s)} &=&   (2\ri)^{s-1} \sum_{k=0}^{s} (-1)^k \binom{s}{k}\non \\
&&\times  \Big\{ \binom{s}{k+1} 
 \pa_{(\a_1 ( \ad_1} \dots \pa_{\a_k \ad_k} D_{\a_{k+1}} \F \,
\bar D_{\ad_{k+1} } \pa_{\a_{k+2} \ad_{k+2} }\dots \pa_{ \a_s) \ad_s  ) } \bar \F \non \\
&& \qquad + 2\ri \binom{s}{k} 
\pa_{(\a_1  (\ad_1} \dots \pa_{\a_k \ad_k}  \F \,
 \pa_{\a_{k+1} \ad_{k+1} }\dots \pa_{ \a_s) \ad_s  ) } \bar \F \Big\}~,
 \label{5.5}
\eea
where one should keep in mind that 
\bea
\binom{s}{s+1} =0~.
\non 
\eea
It is an instructive exercise to check that the conservation equations 
\eqref{5.3} are satisfied.
Choosing $s=1$ in \eqref{5.5} gives the well-known supercurrent 
\cite{FZ}
\bea
J_{\a\ad} = D_\a \F \, \bar D_\ad \bar \F +2\ri (\F\, \pa_{\a\ad} \bar \F 
- \pa_{\a\ad} \F \, \bar \F )~.
\eea
The higher spin supercurrent  \eqref{5.5} may be compared with the 3D $\cN=2$ result 
reported in \cite{NSU}. 

Similar to the bosonic superfield prepotentials  $H_{ \a (s) \ad (s) }$,
one may define  conserved higher spin  current supermultiplets
associated with the fermionic superfield prepotentials $\J_{ \a (s) \ad (s-1) }$.
Consider a source term of the form 
\bea
S^{(s)}_{\rm source}=\int \rd^4x \rd^2 \q  \rd^2 \bar \q \, \J^{ \a (s) \ad (s-1) } 
J_{ \a (s) \ad (s-1) } +{\rm c.c.}
\eea
In order for $S^{(s)}_{\rm source}$ to be invariant under the superconformal
transformations, $J_{\a(s) \ad(s-1)}$ must be
superconformal primary of weight $(1 +\frac{s}{2}, \hf + \frac{s}{2})$. 
In order for $S^{ (s)}_{\rm source}$ to be invariant under the gauge transformations 
 \eqref{2.6},  the superfield $J_{\a(s) \ad(s-1)}$ with  $s>1$ 
 must obey the conservation equations
 \bea
 D^\b J_{\b \a_1 \dots \a_{s-1} \ad_1 \dots \ad_{s-1} }=0 ~,\qquad
 \bar D^\bd J_{\a_1 \dots \a_s \bd \ad_1 \dots \ad_{s-2}}=0~.
 \label{5.9}
 \eea
In the $s=1$ case, the conservation equations are \cite{KT}
\bea
D^\b J_\b =0~, \qquad \bar D^2 J_\a =0~,
\label{5.10}
\eea
as a consequence of \eqref{3.21}.

Conserved current multiplets $J_{\a(s) \ad(s-1)}$,
with $s=1, 2, \dots,$ may be constructed 
from two free massless chiral superfields $\F_+$ and $\F_-$, 
\bea
\bar D_\ad \F_\pm  =0~, \qquad D^2 \F_\pm =0~.
\eea
One may check that the following composite\footnote{SMK is grateful to 
Jessica Hutomo for pointing out a sign error in eq. \eqref{5.12}
in the published version of this work.}
\bea
J_{\a(s) \ad(s-1)}
&=&   (2\ri)^{s-1} \sum_{k=0}^{s-1}(-1)^k  \binom{s-1}{k}\non \\
&&\times  \Bigg\{ \binom{s}{k+1} 
 \pa_{(\a_1  (\ad_1} \dots \pa_{\a_k \ad_k} D_{\a_{s}} \!\F_+ ~
 \pa_{\a_{k+1} \ad_{k+1} }\dots \pa_{ \a_{s-1}) \ad_{s-1}  ) } 
\F_- \non \\
&& \qquad - \binom{s}{k} 
\pa_{(\a_1  (\ad_1} \dots \pa_{\a_k \ad_k}  \F_+ ~
 \pa_{\a_{k+1} \ad_{k+1} }\dots \pa_{ \a_{s-1} \ad_{s-1}  ) } D_{\a_s )} \F_- \Bigg\}~~~~
 \label{5.12}
 \eea
 obeys the conservation equations \eqref{5.9} for $s>1$.
One may also  check that  $J_{\a(s) \ad (s-1)} $ is (anti)symmetric, 
$ J_{\a(s) \ad (s-1)} \to (-1)^s J_{\a(s) \ad (s-1)} $,
with respect to 
the interchange $\F_+ \leftrightarrow \F_-$.
Choosing $s=1$ in \eqref{5.12} gives the composite
\bea
J_\a = \F_- \stackrel{\longleftrightarrow}{D_\a} \F_+ ~,
\label{5.13}
\eea
which obeys the conservation equations \eqref{5.10}. The superfield $J_\a$ 
contains a conserved fermionic current  $j_{\a\b\bd} = j_{\b \a \bd}$
that corresponds to the second supersymmetry current.\footnote{This conserved current may be chosen as 
$j_{\a\b\bd} = \Big\{ \big[ D_{(\b}, \bar D_\bd \big] J_{\a)} 
+\frac{2}{3} \ri \pa_{(\b \bd} J_{\a)}\Big\}\Big|_{\q=0} $. 
 Also associated with $J_\a$ is the off-shell conserved current 
 $v_{\a\b\bd} = \pa_{\b\bd} J_\a
-2 \ve_{\a\b} \pa_{\g\bd} J^\g$
such that
$\pa^{\b\bd} v_{\a\b\bd } =0$ for any $J_\a$.
} 
The conserved current multiplet \eqref{5.13} is obtained 
by reducing the $\cN=2$ supercurrent of a free massless hypermultiplet 
to $\cN=1$ superspace, see \cite{KT} for more details. 

In the case of a free $\cN=2$ hypermultiplet described in terms of two $\cN=1$ 
chiral scalars $\F_+$ and $\F_-$, the action \eqref{5.1} should be replaced with 
\bea
S[\F_\pm, \bar \F_\pm; H, \J, \bar \J]&=&  \int \rd^4x \rd^2 \q  \rd^2 \bar \q \, \Bigg\{ 
\F_+ \bar \F_+ +\F_- \bar \F_-  
+ \sum_{s=1}^{\infty} H^{ \a (s) \ad (s) } J_{ \a (s) \ad (s) }\non \\
&&\qquad \qquad 
+ \sum_{s=1}^{\infty} \Big[\J^{ \a (s) \ad (-1) } J_{ \a (s) \ad (s-1) } +{\rm c.c.}
\Big]\Bigg\}~.
\eea
The current superfields $J_{ \a (s) \ad (s) }$ and $J_{ \a (s) \ad (s-1) }$ are $\cN=1$
 components of a conserved $\cN=2$ supermultiplet. 
\\

\noindent
{\bf Acknowledgements:}\\
SMK is grateful to Arkady Tseytlin and Misha Vasiliev for useful conversations. 
ST acknowledges hospitality of the School of Physics at 
the University of Western Australia in November 2013 when this project was initiated. 
 The work of SMK and ST is supported in part by the
Australian Research Council, project DP140103925.
The work of RM is supported in part by the Alexander von Humboldt Foundation and by the Science Committee of the Ministry of Science and Education of the Republic of Armenia under contract 15T-1C233Ó.


\appendix

\section{The Grimm-Wess-Zumino superspace geometry}\label{AppendixA}

In describing the Grimm-Wess-Zumino superspace geometry \cite{GWZ}, 
we follow the notation and conventions of \cite{BK}.\footnote{These conventions 
are similar to those
of Wess and Bagger \cite{WB}. To convert the notation of \cite{BK} to that of \cite{WB}, one
replaces $R \rightarrow 2 R$, $G_{\alpha \ad} \rightarrow 2 G_{\alpha \ad}$, and
$W_{\alpha \beta \gamma} \rightarrow 2 W_{\alpha \beta \gamma}$.
In addition, the vector derivative has to be changed by the rule 
$\cD_a \to \cD_a +\frac{1}{4}\ve_{abcd}G^b M^{cd}$, where $G_a$ corresponds to \cite{BK}. 
Finally, the spinor Lorentz generators $(\s_{ab})_\a{}^\b$ and 
$({\tilde \s}_{ab})^\ad{}_\bd$  used in \cite{BK} have an extra minus sign as compared with \cite{WB}, 
specifically $\s_{ab} = -\frac{1}{4} (\s_a \tilde{\s}_b - \s_b \tilde{\s}_a)$ and 
 $\tilde{\s}_{ab} = -\frac{1}{4} (\tilde{\s}_a {\s}_b - \tilde{\s}_b {\s}_a)$.  } 
In particular,  the coordinates of $\cN=1$ curved superspace $\cM^{4|4}$ are denoted 
$z^M = (x^m, \q^\m , {\bar \q}_{\dot \m})$.
The superspace geometry is described 
by covariant derivatives of the form
\bea
\cD_A &=& (\cD_a , \cD_\a ,\cDB^\ad ) = E_A + \O_A~.
\eea
Here $E_A$ denotes the inverse vielbein, 
$E_A = E_A{}^M  \pa_M $,
and $\O_A$  the Lorentz connection, 
\bea
\O_A = \hf\,\O_A{}^{bc}  M_{bc}
= \O_A{}^{\b \g} M_{\b \g}
+\O_A{}^{\bd \gd} {\bar M}_{\bd \gd} ~,
\eea
with $M_{bc} =-M_{cb}\Leftrightarrow 
( M_{\b\g}= M_{\g\b} , {\bar M}_{\bd \gd} = \bar M_{\gd \bd})$
the Lorentz generators.
The covariant derivatives obey the following anti-commutation relations:
\begin{subequations}\label{algebra}
\bea
&& {} \qquad \{ \cD_\a , {\bar \cD}_\ad \} = -2{\rm i} \cD_{\a \ad} ~, 
 \\
\{\cD_\a, \cD_\b \} &=& -4{\bar R} M_{\a \b}~, \qquad
\{ {\bar \cD}_\ad, {\bar \cD}_\bd \} =  4R {\bar M}_{\ad \bd}~,  \label{A.3b}  \\
\left[ { \bar \cD}_{\ad} , \cD_{ \b \bd } \right]
     & = & -{\rm i}{\ve}_{\ad \bd}
\Big(R\,\cD_\b + G_\b{}^{\dot{\g}}  \cDB_{\dot{\g}}
-(\cDB^\gd G_\b{}^{\dot{\d}})
{\bar M}_{\gd \dot{\d}}
+2W_\b{}^{\g \d}
M_{\g \d} \Big)
- {\rm i} (\cD_\b R)  {\bar M}_{\ad \bd}~,~~~~~~~ ~~ \\
\left[ \cD_{\a} , \cD_{ \b \bd } \right]
     & = &
     {\rm i}
     {\ve}_{\a \b}
\Big({\bar R}\,\cDB_\bd + G^\g{}_\bd \cD_\g
- (\cD^\g G^\d{}_\bd)  M_{\g \d}
+2{\bar W}_\bd{}^{\gd \dot{\d}}
{\bar M}_{\gd \dot{\d} }  \Big)
+ {\rm i} (\cDB_\bd {\bar R})  M_{\a \b}~,  \\
\big[ \cD_{\a\ad} , \cD_{\b \bd} \big] &=& \ve_{\ad \bd} \j_{\a \b} +\ve_{\a\b} \j_{\ad\bd} ~,
\label{algebra_d}
\eea
where
\bea
\j_{\a\b} &:=& -\ri G_{( \a }{}^\gd \cD_{\b)\gd} +\hf (\cD_{(\a} R) \cD_{\b )} 
+\hf (\cD_{(\a }G_{\b )}{}^\gd )\bar \cD_\gd +W_{\a\b}{}^\g \cD_\g \non \\
&& +\frac{1}{4} \big((\bar \cD^2 -8R) \bar R \big)M_{\a\b} + (\cD_{( \a} W_{\b )}{}^{\g\d } )M_{\g\d}
-\hf (\cD_{( \a }\bar \cD^\gd G_{\b )}{}^\dd)  \bar M_{\gd \dd}~,  \label{algebra_e}\\
\j_{\ad\bd} &:=&- \ri G_{\g (\ad } \cD^\g {}_{\bd )} -\hf (\bar \cD_{(\ad} \bar R) \bar \cD_{\bd )}
-\hf (\bar \cD_{(\ad } G^\g{}_{\bd )} )\cD_\g -\bar W_{\ad \bd }{}^\gd \bar \cD_\gd \non \\
&& +\frac{1}{4}\big( (\cD^2 - 8 \bar R)  R\big) \bar M_{\ad \bd}
-(\bar \cD_{(\ad} \bar W_{\bd)}{}^{\gd \dd} )\bar M_{\gd \dd} 
+ \hf (\bar \cD_{(\ad} \cD^\g G^\d{}_{\bd)} )M_{\g\d}~. 
\label{algebra_f}
\eea 
\end{subequations}
The torsion tensors $R$, $G_a = {\bar G}_a$ and
$W_{\a \b \g} = W_{(\a \b\g)}$ satisfy the Bianchi identities
\begin{subequations}
\bea
\cDB_\ad R&=& 0~, \qquad \cDB_\ad W_{\a \b \g} = 0~, \label{2.4a} \\
\cDB^\gd G_{\a \gd} &=& \cD_\a R~, \label{2.4b} \\
\cD^\g W_{\a \b \g} &=& {\rm i} \,\cD_{(\a }{}^\gd G_{\b) \gd}~.
\label{2.4c} 
\eea
\end{subequations}

A supergravity gauge transformation is defined to act on the covariant derivatives
and any tensor superfield $U$ (with its indices suppressed) by the rule
\begin{subequations}\label{A.5}
\bea
\d_\cK \cD_A = [\cK, \cD_A] ~, \qquad \d_\cK U = \cK U~, 
\eea
where the gauge parameter $\cK$ has the explicit form 
\bea
\cK = \x^B \cD_B + \hf K^{bc} M_{bc} 
= \x^B \cD_B + K^{\g\d} M_{\g \d} + \bar K^{  \gd \dd} \bar M_{ \gd \dd} = \bar \cK
\label{K}
\eea
\end{subequations}
and describes a general coordinate transformation generated 
by the supervector field $\x=\x^B E_B$ as well as a local Lorentz transformation generated 
by the antisymmetric tensor $K^{bc} $.

\section{Super-Weyl transformations}\label{AppendixB}

The algebra of covariant derivatives \eqref{algebra} 
preserves its functional form
under super-Weyl transformations \cite{HT} 
\begin{subequations} 
\label{superweyl}
\bea
\d_\s \cD_\a &=& ( {\bar \s} - \hf \s)  \cD_\a + (\cD^\b \s) \, M_{\a \b}  ~, \\
\d_\s \bar \cD_\ad & = & (  \s -  \hf {\bar \s})
\bar \cD_\ad +  ( \bar \cD^\bd  {\bar \s} )  {\bar M}_{\ad \bd} ~,\\
\d_\s \cD_{\a\ad} &=& \hf( \s +\bar \s) \cD_{\a\ad} 
+\frac{\ri}{2} (\bar \cD_\ad \bar \s) \cD_\a + \frac{\ri}{2} ( \cD_\a  \s) \bar \cD_\ad \non \\
&& + (\cD^\b{}_\ad \s) M_{\a\b} + (\cD_\a{}^\bd \bar \s) \bar M_{\ad \bd}~,
\eea
\end{subequations}
where $\s$ is an arbitrary covariantly chiral scalar superfield,  $\bar \cD_\ad \s =0$. 
The torsion tensors in  \eqref{algebra} transform
 as follows:
\begin{subequations} 
\bea
\d_\s R &=& 2\s R +\frac{1}{4} (\bar \cD^2 -4R ) \bar \s ~, \\
\d_\s G_{\a\ad} &=& \hf (\s +\bar \s) G_{\a\ad} +\ri \cD_{\a\ad} ( \s- \bar \s) ~, 
\label{s-WeylG}\\
\d_\s W_{\a\b\g} &=&\frac{3}{2} \s W_{\a\b\g}~.
\label{s-WeylW}
\eea
\end{subequations} 
The local transformations \eqref{A.5} and \eqref{superweyl} constitute the gauge freedom of conformal supergravity.


\begin{footnotesize}

\end{footnotesize}


\begin{thebibliography}{66}



\bibitem{Fronsdal1}
  C.~Fronsdal,
  ``Massless fields with integer spin,''
  Phys.\ Rev.\  D {\bf18},   3624 (1978);
%
  J.~Fang and C.~Fronsdal,
  ``Massless fields with half-integral spin,''
  Phys.\ Rev.\  D {\bf 18}, 3630 (1978).


\bibitem{Fronsdal2}
  C.~Fronsdal,
  ``Singletons and massless, integral-spin fields on de Sitter space,''
  Phys.\ Rev.\  D {\bf 20},  848 (1979);
%
  J.~Fang and C.~Fronsdal,
  ``Massless, half-integer-spin fields in de Sitter space,''
  Phys.\ Rev.\  D {\bf 22},  1361 (1980).


\bibitem{FV} 
    E.~S.~Fradkin and M.~A.~Vasiliev,
  ``On the gravitational interaction of massless higher spin fields,''
  Phys.\ Lett.\ B {\bf 189}, 89 (1987);
  ``Cubic interaction in extended theories of massless higher spin fields,''
  Nucl.\ Phys.\ B {\bf 291}, 141 (1987).

\bibitem{Vasiliev} 
  M.~A.~Vasiliev,
  ``Consistent equation for interacting gauge fields of all spins in (3+1)-dimensions,''
  Phys.\ Lett.\ B {\bf 243}, 378 (1990).

\bibitem{FT} 
E.~S.~Fradkin and A.~A.~Tseytlin,  ``Conformal supergravity,''
Phys.\ Rept.\  {\bf 119}, 233 (1985).

\bibitem{FL} 
  E.~S.~Fradkin and V.~Y.~Linetsky,
  ``Superconformal higher spin theory in the cubic approximation,''
  Nucl.\ Phys.\ B {\bf 350}, 274 (1991).

\bibitem{BK} I.~L.~Buchbinder and S.~M.~Kuzenko,
{\it Ideas and Methods of Supersymmetry and Supergravity or a Walk Through Superspace},
IOP, Bristol, 1995 (Revised Edition 1998).


\bibitem{GWZ} 
  R.~Grimm, J.~Wess and B.~Zumino,
  ``Consistency checks on the superspace formulation of supergravity,''
  Phys.\ Lett.\ B {\bf 73}, 415 (1978);
  ``A complete solution of the Bianchi identities in superspace,''
 Nucl.\ Phys.\ B {\bf 152}, 255 (1979).


\bibitem{HT}
P.~S.~Howe and R.~W.~Tucker,
``Scale invariance in superspace,''
Phys.\ Lett.\ B {\bf 80}, 138 (1978).


\bibitem{WZ}
J.~Wess and B.~Zumino,
 ``Superfield Lagrangian for supergravity,''
 Phys.\ Lett.\  B {\bf 74}, 51 (1978).


\bibitem{old}
K.~S.~Stelle and P.~C.~West,
``Minimal auxiliary fields for supergravity,''
Phys.\ Lett.\  B {\bf 74},  330 (1978);
S.~Ferrara and P.~van Nieuwenhuizen,
``The auxiliary fields of supergravity,''
Phys.\ Lett.\  B {\bf 74}, 333 (1978).


\bibitem{Howe}
P.~S.~Howe,
``A superspace approach to extended conformal supergravity,''
  Phys.\ Lett.\ B {\bf 100}, 389 (1981);
``Supergravity in superspace,''  Nucl.\ Phys.\  B {\bf 199}, 309 (1982).
  
\bibitem{Butter}
D.~Butter,  ``N=1 conformal superspace in four dimensions,''
Annals Phys.\  {\bf 325}, 1026 (2010)
  [arXiv:0906.4399 [hep-th]].


\bibitem{KPS}
S.~M.~Kuzenko,  V.~V.~Postnikov and A.~G.~Sibiryakov,
``Massless gauge superfields of higher half-integer superspins,''
JETP Lett.\  {\bf 57},    534 (1993) 
[Pisma Zh.\ Eksp.\ Teor.\ Fiz.\  {\bf 57},  521 (1993)].
  
\bibitem{KS93}
S.~M.~Kuzenko and A.~G.~Sibiryakov,
``Massless gauge superfields of higher integer superspins,''
JETP Lett.\  {\bf 57},   539 (1993)  
[Pisma Zh.\ Eksp.\ Teor.\ Fiz.\  {\bf 57}, 526 (1993)].



\bibitem{FZ2}
S.~Ferrara and B.~Zumino,
``Structure of conformal supergravity,''  Nucl.\ Phys.\  B {\bf 134}, 301 (1978).


\bibitem{OS}
V.~Ogievetsky and E.~Sokatchev,
``On vector superfield generated by supercurrent,''
Nucl.\ Phys.\  B {\bf 124}, 309 (1977).


\bibitem{Siegel}
W.~Siegel, ``A derivation of the supercurrent superfield,''
Harvard  preprint  HUTP-77/A089 (December, 1977). 

\bibitem{KU} 
  T.~Kugo and S.~Uehara,
  ``$N=1$ superconformal tensor calculus: Multiplets with external Lorentz indices and spinor derivative operators,''
  Prog.\ Theor.\ Phys.\  {\bf 73}, 235 (1985).

\bibitem{Park}
  J.~H.~Park,
 ``Superconformal symmetry and correlation functions,''
  Nucl.\ Phys.\ B {\bf 559}, 455 (1999)
   [hep-th/9903230].


\bibitem{GGRS}
S.~J.~Gates Jr., M.~T.~Grisaru, M.~Ro\v{c}ek and W.~Siegel,
{\it Superspace, or One Thousand and One Lessons in Supersymmetry},
Benjamin/Cummings (Reading, MA),  1983, hep-th/0108200.

\bibitem{GS}
S.~J.~Gates  Jr. and W.~Siegel,
``(3/2, 1) superfield of O(2) supergravity,''  Nucl.\ Phys.\  {\bf B164}, 484 (1980).

 
\bibitem{Keck}
  B.~W.~Keck,
 ``An alternative class of supersymmetries,''
J.\ Phys.\ A  {\bf 8}, 1819 (1975).

\bibitem{Zumino77}
B.~Zumino, ``Nonlinear realization of supersymmetry in de Sitter space,''
Nucl.\ Phys.\  B {\bf 127}, 189 (1977).

\bibitem{IS}
E.~A.~Ivanov and A.~S.~Sorin,
``Superfield formulation of OSp(1,4) supersymmetry,''
J.\ Phys.\ A  {\bf 13}, 1159 (1980).

\bibitem{KS94}
S.~M.~Kuzenko and A.~G.~Sibiryakov,
``Free massless higher-superspin superfields on the anti-de Sitter superspace"
Phys.\ Atom.\ Nucl.\  {\bf 57}, 1257 (1994) 
 [Yad.\ Fiz.\  {\bf 57}, 1326 (1994)] [arXiv:1112.4612 [hep-th]].

\bibitem{BK88}
  I.~L.~Buchbinder and S.~M.~Kuzenko,
 ``Quantization of the classically equivalent theories in the superspace of simple supergravity and quantum equivalence,''
  Nucl.\ Phys.\ B {\bf 308}, 162 (1988). 

\bibitem{Siegel78}
W.~Siegel,
``Solution to constraints in Wess-Zumino supergravity formalism,''
Nucl.\ Phys.\  B {\bf 142}, 301 (1978). 

\bibitem{Zumino} 
  B.~Zumino,
  ``Supergravity and superspace,''
in {\it Recent Developments in  Gravitation - Carg\`ese 1978}, 
M. L\'evy and S. Deser (Eds.), N.Y., Plenum Press, 1979, pp. 405--459.


\bibitem{GrisaruSiegel} 
  M.~T.~Grisaru and W.~Siegel,
 ``Supergraphity (I). Background field formalism,''
  Nucl.\ Phys.\ B {\bf 187}, 149 (1981);
 ``Supergraphity (II). Manifestly covariant rules and higher loop finiteness,''
Nucl.\ Phys.\ B {\bf 201}, 292 (1982)

\bibitem{KTs} 
  S.~M.~Kuzenko,
  ``Higher spin super-Cotton tensors and generalisations of the linearÐchiral duality in three dimensions,''
  Phys.\ Lett.\ B {\bf 763}, 308 (2016)
  [arXiv:1606.08624 [hep-th]];
S.~M.~Kuzenko and M.~Tsulaia,
  ``Off-shell massive N=1 supermultiplets in three dimensions,''
  Nucl.\ Phys.\ B {\bf 914}, 160 (2017)
  [arXiv:1609.06910 [hep-th]].

\bibitem{KO} 
S.~M.~Kuzenko and D.~X.~Ogburn,
  ``Off-shell higher spin N=2 supermultiplets in three dimensions,''
  Phys.\ Rev.\ D {\bf 94}, no. 10, 106010 (2016)
  [arXiv:1603.04668 [hep-th]].

\bibitem{Tseytlin2002} 
  A.~A.~Tseytlin,
  ``On limits of superstring in AdS(5) x S**5,''
  Theor.\ Math.\ Phys.\  {\bf 133}, 1376 (2002)
  [Teor.\ Mat.\ Fiz.\  {\bf 133}, 69 (2002)]
  [hep-th/0201112].
  
 \bibitem{Segal} 
  A.~Y.~Segal,
  ``Conformal higher spin theory,''
  Nucl.\ Phys.\ B {\bf 664}, 59 (2003)
  [hep-th/0207212]. 

\bibitem{BJM} 
  X.~Bekaert, E.~Joung and J.~Mourad,
  ``Effective action in a higher-spin background,''
  JHEP {\bf 1102}, 048 (2011)
  [arXiv:1012.2103 [hep-th]].

\bibitem{BJM2} 
  X.~Bekaert, E.~Joung and J.~Mourad,
  ``On higher spin interactions with matter,''
  JHEP {\bf 0905}, 126 (2009)
  [arXiv:0903.3338 [hep-th]].

\bibitem{FZ}
  S.~Ferrara and B.~Zumino,
``Transformation properties of the supercurrent,''
  Nucl.\ Phys.\  B {\bf 87}, 207 (1975).

\bibitem{HST}
P.~S.~Howe, K.~S.~Stelle and P.~K.~Townsend,
``Supercurrents,''  Nucl.\ Phys.\  B {\bf 192}, 332 (1981).

\bibitem{Siegel:1980bp} 
  W.~Siegel,
  ``On-shell O($N$) supergravity in superspace,''
  Nucl.\ Phys.\ B {\bf 177}, 325 (1981).


\bibitem{CDT} 
N.~S.~Craigie, V.~K.~Dobrev and I.~T.~Todorov,
``Conformally covariant composite operators in quantum chromodynamics,''
  Annals Phys.\  {\bf 159}, 411 (1985).

\bibitem{NSU}
  A.~A.~Nizami, T.~Sharma and V.~Umesh,
  ``Superspace formulation and correlation functions of 3d superconformal field theories,''
  JHEP {\bf 1407}, 022 (2014)
  [arXiv:1308.4778 [hep-th]].


\bibitem{KT}
S.~M.~Kuzenko and S.~Theisen,
``Correlation functions of conserved currents in N = 2 superconformal
theory,''  Class.\ Quant.\ Grav.\  {\bf 17}, 665 (2000)  [hep-th/9907107]. 

\bibitem{WB} J.~Wess and J.~Bagger,
{\it Supersymmetry and Supergravity},
Princeton University Press, Princeton, 1983 (Second Edition 1992).

\end{thebibliography}
\end{document}